# Profiling of FSHR Negative Allosteric Modulators on LH/CGR Reveals Biased Antagonism with Implications in Steroidogenesis

Mohammed Akli Ayoub[1,2,*], Romain Yvinec[1], Gwenhaël Jégot[1], James A. Dias[3], Sonia-Maria Poli[4], Anne Poupon[1], Pascale Crépieux[1], and Eric Reiter[1]

[1] PRC, INRA, CNRS, IFCE, Université de Tours, 37380, Nouzilly, France.

[2] LE STUDIUM® Loire Valley Institute for Advanced Studies, 45000, Orléans, France.

[3] Department of Biomedical Sciences, State University of New York at Albany, Albany, New York.

[4] Addex Pharma S.A, Plan-les-Ouates, Geneva, Switzerland.

**\* Correspondence**: Institut National de la Recherche Agronomique (INRA) UMR85, CNRS-Université François-Rabelais UMR7247, Physiologie de la Reproduction et des Comportements - Nouzilly 37380, France - Email: *Mohammed.Ayoub@tours.inra.fr*

**Abstract**

Biased signaling has recently emerged as an interesting means to modulate the function of many G protein-coupled receptors (GPCRs). Previous studies reported two negative allosteric modulators (NAMs) of follicle-stimulating hormone receptor (FSHR), ADX68692 and ADX68693, with differential effects on FSHR-mediated steroidogenesis and ovulation. In this study, we attempted to pharmacologically profile these NAMs on the closely related luteinizing hormone/chorionic gonadotropin hormone receptor (LH/CGR) with regards to its canonical Gs/cAMP pathway as well as to ß-arrestin recruitment in HEK293 cells. The NAMs' effects on cAMP, progesterone and testosterone production were also assessed in murine Leydig tumor cell line (mLTC-1) as well as primary rat Leydig cells. We found that both NAMs strongly antagonized LH/CGR signaling in the different cell models used with ADX68693 more potent than ADX68692 to inhibit hCG-induced cAMP production in HEK293, mLTC-1 and Leydig cells as well as ß-arrestin 2 recruitment in HEK293 cells. Interestingly, differential antagonism of the two NAMs on hCG-promoted steroidogenesis in mLTC-1 and Leydig cells was observed, eventhough both NAMs inhibited cAMP pathways. Indeed, while a significant inhibition of testosterone production by the two NAMs was observed in both cell types, progesterone production was only inhibited by ADX68693 in primary rat Leydig cells. In addition, while ADX68693 totally abolished testosterone production at 10 μM, ADX68692 had only a partial effect in both mLTC-1 and primary rat Leydig cells. These observations suggest biased effects of the two NAMs on LH/CGR-dependent pathways controlling steroidogenesis. Interestingly, the pharmacological profiles of the two NAMs with respect to steroidogenesis were found to differ from that previously shown on FSHR. This illustrates the complexity of signaling pathways controlling FSHR- and LH/CGR-mediated steroidogenesis, suggesting differential implication of cAMP and ß-arrestins mediated by FSHR and LH/CGR. Together, our data demonstrate that ADX68692 and ADX68693 are biased NAMs at the LH/CGR in addition to the FSHR. These pharmacological characteristics are important to consider for potential contraceptive and therapeutic applications based on such compounds.

# 1. Introduction

The follicle-stimulating hormone receptor (FSHR) and luteinizing hormone/chorionic gonadotropin hormone receptor (LH/CGR) are G protein-coupled receptor (GPCR) members known for their central role in the control of reproduction. Thus, a particular interest has been given to these two receptors with regards to infertility, contraception, estrogen-dependent diseases and other disorders of the reproductive system in medicine and in animal husbandry. One of the most important aspects in the recent advances on FSHR and LH/CGR is related to the development of small molecules to positively or negatively modulate their activity with the aim to better understand their mechanism of activation and to develop potential therapeutics. These agents may act as orthosteric ligands at the binding site or as allosteric modulators. Indeed, many studies reported small molecules as potential tools to study underlying mechanisms that enable successful reproduction . In addition, small molecules acting on FSHR and LH/CGR have been proposed as alternative oral therapeutics for infertility treatment (agonists) or contraception strategies (antagonists) . Small molecule agonists of LH/CGR were reported to efficiently induce ovulation . Moreover, dimeric molecules were developed with dual effects as antagonist on FSHR and agonist on LH/CGR . For FSHR, recent studies reported interesting small molecules acting either as antagonists , negative allosteric modulators (NAMs) or as positive allosteric modulators (PAMs) or agonists . Indeed, a thiazolidinone derivative has been reported to activate FSHR signaling in CHO cells and estradiol production in cultured rat granulosa cells . Optimization of substituted benzamides led to more FSHR-selective molecules relative to other closely related GPCRs, such as LH/CGR and thyroid stimulating hormone receptor (TSHR) with better pharmacokinetic properties .

The initial FSHR NAM molecule reported was ADX61623, which blocked FSHR-mediated cAMP as well as progesterone but not estradiol production in rat granulosa primary cells . However, ADX61623 did not affect FSH-induced preovulatory follicle development, limiting its application as a nonsteroidal contraceptive . Two other NAMs, ADX68692 and ADX68693, with structural similarities to ADX61623, were tested and exhibited different antagonistic profile on FSHR in rat granulosa primary cells . In fact, while ADX68692 blocked FSHR-promoted cAMP, progesterone and estradiol production, ADX68693 inhibited cAMP and progesterone with the same efficacy as ADX68692 but did not block estradiol production . This study proposed a potential application of ADX68692 as a nonsteroidal contraceptive since it was also orally active in blocking FSH-induced follicular growth. Based on the functional difference between both analogs, it appeared that the contraceptive effects require the blockade of the production of both progesterone and estradiol. Both FSHR and LH/CGR are involved in the control of steroid sex hormones, are co-expressed in granulosa cells at specific stages , belong to leucine-rich repeat sub-family of GPCRs and are known to couple to the canonical Gs/cAMP/PKA signaling pathway . Because of the important structural similarity between FSHR and LH/CGR at the level of their transmembrane domains, one can hypothesize that ADX68692 and ADX68693 may also modulate LH/CGR activity thereby expanding the spectra of their pharmacological actions. Moreover, previous studies revealed biased inhibitory profiles of ADX68692 *versus* ADX68693 on FSHR-mediated steroidogenesis, albeit the molecular mechanisms underneath remained unknown. Noteworthy, FSHR and LH/CGR have been reported to be susceptible to biased activation with implication of the non-canonical ß-arrestin-dependent signaling pathway . In this context, it is still unclear whether the two ADX compounds may have biased antagonistic effects involving preferential actions on G proteins *versus* ß-arrestins.

In this study, we pharmacologically profiled the two compounds, ADX68692 and ADX68693, on LH/CGR. For this, we investigated their effects on the canonical Gs/cAMP pathway as well as on ß-arrestin 2 recruitment in HEK293 cells using BRET technology as previously described . Moreover, the action of these NAMs was examined in mLTC-1 as well as primary rat Leydig cells known to endogenously express LHR , by assessing their effects on hCG-promoted cAMP, progesterone and testosterone production. Finally, to further explore the differential effects of the two NAMs with respect to the different hCG-promoted responses in mLTC-1 and primary rat Leydig cells bias factors were calculated.

## 2. Materials and Methods

### 2.1. Materials and plasmids

The FSHR NAMs ADX68692 (MW 401.44) and ADX68693 (MW 352.42) were prepared by Addex Pharmaceuticals S.A (Geneva, Switzerland). The specificity of each of the FSHR NAMs was reported previously . The plasmid encoding human FSHR was generated as previously described and hLHR plasmid was obtained from A. Ulloa-Aguirre (Universidad Nacional Autónoma de México, México, Mexico). The other plasmids encoding the different BRET/FRET sensors and fusion proteins were generously provided as follows: Rluc8-fused hLH/CGR from A. Hanyaloglu (Imperial College, London, UK), yPET-ß-arrestin 2 from M.G. Scott (Cochin Institute, Paris, France), CAMYEL from L.I. Jiang (University of Texas, Texas, USA), V2R-Rluc8 from K.D. Pfleger (Harry Perkins Institute of Medical Research, Perth, Australia). Recombinant hFSH was kindly gifted by Merck-Serono (Darmstadt, Germany), hCG was kindly donated by Y. Combarnous (CNRS, Nouzilly, France), and desmopressin (DDAVP) a synthetic form of vasopressin was purchased from Sigma-Aldrich (St. Louis, MO, USA). Ninety-six-and 384-well white microplates were from Greiner Bio-One (Courtaboeuf, France). Coelenterazine h substrate was purchased from Interchim (Montluçon, France).

### 2.2. Cell culture and transfection

HEK293 cells were grown in complete medium (DMEM supplemented with 10% (v/v) fetal bovine serum, 4.5 g/l glucose, 100 U/ml penicillin, 0.1 mg/ml streptomycin, and 1 mM glutamine, all from Invitrogen, Carlsbad, CA). Transient transfections were performed in 96-well plates using Metafectene PRO (Biontex, München, Germany) following the manufacturer's protocol. Briefly, for each well the different combinations of coding plasmids were used as follows: 200 ng of total plasmid per well were resuspended in 25 μl of serum-free DMEM and mixed with Metafectene PRO (0.5 μl/well) previously preincubated 5 minutes at room temperature in 25 μl serum-free DMEM (2x25 μl/well). Then the two solutions of serum-free DMEM containing plasmids and Metafectene

PRO were mixed and incubated for 20 minutes at room temperature. Cells ($10^5$ in 200 μl/well) were then incubated with the final plasmid-Metafectene PRO mix (50 μl/well) and cultured in DMEM supplemented with 10% fetal bovine serum for forty-eight hours before experiments.

*2.2. Leydig cells isolation*

Leydig cells were isolated from the testes of mature 52-day-old Wistar rats as previously described (Matinat et al, 2005).

*2.3. BRET measurements*

Forty-eight hours after transfection, cells were washed with PBS and BRET measurements were performed depending on the experiments, as described previously . For the endpoint dose-response analyses, cells were first preincubated for 20 minutes at 37°C in the presence or absence of increasing concentrations of ADX68692 and ADX68693 in 30 μl/well of PBS 1X, HEPES 5 mM. Then, 10 μl/well of increasing doses of 4X hCG in PBS 1X, HEPES 5 mM were added and cells were incubated for 30 minutes at 37°C as indicated. Then BRET measurements were performed upon addition of 10 μl/well of 5X (5 μM final concentration) coelenterazine h in PBS 1X, HEPES 5 mM using a Mithras LB 943 plate reader. For the real-time BRET kinetics, cells were first resuspended in 30 μl/well of PBS 1X, HEPES 10 mM containing or not 10 μM of ADX68692 and ADX68693 and then BRET measurements were immediately performed upon addition of 10 μl/well of coelenterazine h (5 μM final concentration) and 10 μl/well of hCG (5-fold concentrated).

*2.4. cAMP accumulation measured by Glosensor™ assay*

For the measurement of cAMP accumulation, we also used the Glosensor™ cAMP assay (Promega). Growing cells in 96-well plates (80,000 cells per well) in growing medium were incubated overnight at 37°C. On the following day, culture medium was removed and replaced by 100 μl of equilibration medium per well (DMEM-serum free medium with 4% v/v of GloSensor™ cAMP reagent stock solution) and cells were incubated for 2 hours at room temperature. Then, cells were incubated for 20 minutes in the presence or absence of increasing concentrations of FSHR NAMs before stimulation with FSH (1.3 nM) and rapid measurement of luciferase signal was peformed using a POLARstar OPTIMA luminometer (BMG Labtech, Ortenberg, Germany).

*2.5. cAMP accumulation measured by HTRF®*

Intracellular cAMP levels were measured using a homogeneous time-resolved fluorescence (HTRF®) cAMP dynamic assay kit (CisBio Bioassays, Bagnol sur Cèze, France). Cells were detached and first resuspended in PBS 1X, 10 mM HEPES, 0.1% BSA containing or 10 µM of ADX68692 or ADX68693. After incubation for 20 minutes at 37°C, 5 µl/well (~5000 cells) of cells were seeded into white 384-well microplate and supplemented with 5 µl/well of the stimulation buffer in the absence or presence of hCG as indicated. Cells were then incubated for 30 minutes at 37°C before lysis by addition of 10 µl/well of the supplied conjugate lysis buffer containing d2-labeled cAMP and Europium cryptate-labeled anti-cAMP antibody, both reconstituted according to the manufacturer's instructions. The plate was incubated for 1 h in the dark at room temperature and fluorescence was measured at 620 nm and 665 nm respectively, 50 ms after excitation at 320 nm using a Mithras LB 943 plate reader (Berthold Technologies GmbH & Co. Wildbad, Germany).

*2.6. cAMP reporter gene assay*

HEK293 cells were transiently co-transfected with plasmids encoding the FSHR or LH/CGR and pSOM-Luc coding for a cAMP-sensitive reporter gene, as previously reported . After overnight starvation in DMEM-serum free, cells were first pretreated or not for 1 hour at 37°C with 10 µM of ADX68692 and ADX68693 in DMEM-serum free. Then, cells were stimulated for 6 hours at 37°C with increasing concentrations of gonadotropins in a final volume of 50 µl/well DMEM-serum free. Luciferase luminescence was measured on a Mithras LB 943 plate reader upon addition of 50 µl/well of Bright-Glo™ luciferase substrate in the supplied lysis buffer and incubation of cells for 2-5 minutes at room temperature.

*2.7. Progesterone and testosterone production*

The production of progesterone and testosterone in mLTC-1 and freshly isolated rat Leydig cells was assessed using a homogenous time–resolved fluorescence (HTRF®)-based assay (CisBio Bioassays, Codolet, France). mLTC-1 cells were first cultured in 96-well microplates and starved overnight in serum-free RPMI-1640 medium containing 25 mM HEPES and 0.3 g/L of L-glutamine. The primary rat Leydig cells were resuspenend and seeded in 96-well plate to ~50000 cells/well in 50 μL/well. Cells were pre-treated for 1 hour at 37°C with 10 μM of ADX68692 or ADX68693 prepared in 50 μL/well in serum-free RPMI-1640 medium containing 25 mM HEPES and 0.3 g/L of L-glutamine, then stimulated with increasing concentrations of hCG by adding 10 μL/well of hCG prepared 6X in serum-free RPMI-1640 medium containing 25 mM HEPES and 0.3 g/L of L-glutamine. After incubation 24 hours for mLTC-1 cells and 3 hours for freshly isolated rat Leydig cells at 37°C, 10 μL of the culture supernatant were transferred to a white 384-well microplate and mixed with 10 μL/well of either progesterone or testosterone. HTRF® reagents (5 μL of Europium cryptate-conjugated anti-progesterone/testosterone antibody + 5 μL of d2-conjugated progesterone/testosterone) previously resupended in the supplied conjugate lysis buffer. The 384-well microplate was then incubated for 1 hour at room temperature in the dark before fluorescence was measured at 620 nm and 665 nm respectively, 50 ms after excitation at 320 nm using a Mithras LB 943 plate reader (Berthold Technologies GmbH & Co. Wildbad, Germany).

*2.8. Data and statistical analysis*

Data are presented as "% of response" by taking as 100% the maximal responses of the hormone at 100 nM measured in cells treated with hCG + DMSO in the different assays (dotted lines in the dose curves). The kinetic curves and the sigmoidal dose-responses curves were fitted using Prism 5 graphing software (GraphPad, La Jolla, CA, USA). Two-way ANOVA analysis using Turkey's multiple comparisons test was used to determine statistically significant differences between the different conditions. *** $p\text{-value} < 0.001$, ** $p\text{-value} < 0.01$, * $p\text{-value} < 0.05$.

*2.9. Bias calculations*

Transduction coefficients (R= τ /Ka) and bias factor (B.F.) were obtained after statistical fitting of the operational model for each dose-response curve. Specifically, we followed a similar procedure as in [van der Westhuizen, 2014]. For a given hCG-induced response, in the absence (DMSO) or presence of ADX68692 or ADX68693, the dose-response curves were fitted using the operational model given by equation (1):

$$E = Basal + (Emax-Basal) * (\tau [A])^n / ((\tau [A])^n+(Ka + [A])^n), \quad (1)$$

where [A] denotes the concentration of hCG and E is the quantification of its effect (cAMP and steroid production). The basal parameter is the baseline of the response, Emax is the maximal possible response of the system, τ is the efficacy, Ka is the functional equilibrium dissociation constant of the agonist and n is the Hill slope of the transducer function that links occupancy to response. As detailed in Van der Westhuizen et al. , such an equation (1) is badly parametrized to yield proper identification of the transduction coefficient, so we transformed the equation (1) using standard algebraic manipulations, into equation (2):

$$E = Basal + (Emax-Basal) / (1 + ((1+[A]/10^{\wedge}(\log (Ka))/(10^{\wedge}Log (R) * [A]))^n, \quad (2)$$

Also, for a given response Basal, Emax, and n were imposed to have the same value for all input (response specific parameters) and n was set to 1 to improve parameter identifiability. Finally, τ and Ka were taken as treatment specific. Then, log (R_DMSO), log (R_ADX68692), log (R_ADX68693), log (Ka_ADX68692), and log (Ka_ADX68693) were determined by statistical fitting together with Basal, Emax and n. Data-fitting of such a model were performed using the D2D-software, on Matlab 14, and parameter uncertainty was quantified using the profile likelihood method and standard hessian calculations. For DMSO input (in the absence of NAMs), the log (Ka) was fixed to 0. For all dose-response curves, except testosterone response in the presence of ADX68693 treatment, the transduction coefficient log (R) = log (τ/ Ka) was found to be practically identifiable, with finite narrow confidence interval (of total length from half a log to one log). Mean transduction coefficient together with the standard error on the mean was then estimated. Still for a single dose-response curve, the relative effectiveness between two treatments t1 and t2 (for instance, DMSO vs NAMs) is given by:

**Δlog (τ/ Ka) = log (τ/ Ka)_t1 – log (τ/ Ka)_t2**

To obtain bias factor comparison for two different responses (1 and 2), treatment biases is then given by:

**ΔΔlog (τ/ Ka) = Δlog (τ/ Ka)_respone 1 – Δlog (τ/ Ka)_response 2**

and the bias factor is the exponential of the latter,

**BF= 10^ΔΔlog (τ/ Ka)**

Finally, standard deviation on ligand biases is determined by the standard deviation of the relative effectiveness, the latter being determined by the data-fitting procedure. After standard calculation, mean bias factor together with their standard error were evaluated from the data. Finally, two-way unpaired t-test was performed to obtain the significance of the bias factors.

## 3. Results

In this study we investigated the effects of two compounds, ADX68692 and ADX68693 previously shown acting as NAMs on FSHR *in vitro* and *in vivo* , on LH/CGR. For this, we used three different cell types, HEK293 transiently co-expressing LH/CGR with the different sensors for cAMP and ß-arrestin 2 measurements using bioluminescence resonance energy transfer (BRET) approach and murine Leydig tumor cell line (mLTC-1) as well as primary rat Leydig cells endougenously expressing LHR for the effects on cAMP production and steroidogenesis.

### 3.1. Effects of ADX compounds on LH/CGR-mediated cAMP production

First, we confirmed the antagonistic effect of ADX68692 and ADX68693 on FSHR-promoted cAMP production in transfected HEK293 cells . Using the Glosensor™ cAMP assay, we clearly demonstrated a dose-dependent inhibition of FSH-induced cAMP production in HEK293 cells stably expressing FSHR, with an $IC_{50}$ values around μM (Log $IC_{50}$ values of -5.71 ± 0.21 and -5.12 ± 0.23 for ADX68692 and ADX68693, respectively)(**Fig. 1A**). Next, we examined the effect of the two compounds on LH/CGR-mediated cAMP production in HEK293 cells transiently expressing LH/CGR. Accordingly, BRET-based sensor assay previously reported was conducted with different doses of NAMs combined with increasing doses of hCG as indicated. As shown in **Fig. 1B** and **C**, the dose-dependent response of hCG on its receptor is consistent with a previous study using similar assay . Similarly to FSHR, both ADX68692 (**Fig. 1B**) and ADX68693 (**Fig. 1C**) inhibited LH/CGR-mediated cAMP production in dose-dependent manner, with a significant shift in hCG potency (**Table 1**). However, such an inhibition was not observed at high concentrations of hCG (i.e. 10 and 100 nM). When both compounds were compared at sub-saturating concentration of the hormone (0.1 nM), ADX68693 appeared significantly more potent than ADX68692 at 1 μM and 10 μM (*p-value < 0.05*) on LH/CGR (**Fig. 1D**) with $pIC_{50}$ values of 4.85 ± 0.36 and 5.85 ± 0.14 for ADX68692 and ADX68693, respectively. Finally, in order to demonstrate the specificity of both compounds on gonadotropin receptors, we tested another Gs-coupled receptor, the vasopressin 2 receptor (V2R). As shown in **Fig. 1E**, neither ADX68692 nor ADX68693 (10 μM) affected V2R-mediated cAMP production assessed with similar BRET-based sensor assay upon its activation with

DDAVP. Together, our results confirm the previous data showing ADX68692 and ADX68693 as FSHR NAMs in rat granulosa primary cells and clearly demonstrate that these compounds also inhibit Gs/cAMP pathway mediated by LH/CGR in HEK293 cells.

Next, real-time kinetics of cAMP production were performed using 20 minutes pretreatment with 10 μM of NAMs followed by rapid activation with 1 nM of hCG. As shown in **Fig. 2A**, hCG nicely induced cAMP production in time-dependent manner and both NAMs blocked such responses to different extent. Interestingly, when comparing both NAMs, ADX68693 appeared significantly stronger than ADX68692 on LH/CGR-mediated response, which is consistent with the difference in the inhibitory effect at 1 nM of hCG shown in **Fig. 1B** (for ADX68692) and **C** (for ADX68693). In order to examine whether the NAMs could reverse the gonadotropin-mediated response, we performed real-time kinetics where cells were first stimulated with hCG and either NAM (10 μM) being applied 5 minutes later. This resulted in rapid time-dependent decrease in hCG-promoted cAMP production (**Fig. 2B**), demonstrating the antagonistic action of the two NAMs even when the receptor is fully activated by hCG such as in the physiological situation.

To link our data on Gs/cAMP pathway with an integrated cellular response, a cAMP-responsive element (CRE)-driven luciferase reporter assay was implemented. This assay uses a pSOM-Luc plasmid coding for the firefly luciferase under the control of the CRE of the somatostatin 5' regulatory region, as previously reported . The gonadotropin-promoted luciferase expression was assessed in HEK293 cells transiently co-expressing either FSHR used here as a positive control (**Fig. 3A**) or LH/CGR (**Fig. 3B**) stimulated with increasing concentrations of FSH or hCG, respectively. The treatment of cells with 10 μM of ADX68692 or ADX68693 significantly diminished the maximal responses ($E_{max}$) with a slight effect on the potency of FSH (**Fig. 3A**) and hCG (**Fig. 3B**) on their respective receptors (**Table 1**). These data indicated a negative allosteric effect of the compounds on FSHR as well as LH/CGR. Together with the cAMP data, these data demonstrate the antagonism of both NAMs on the cAMP/PKA/CREB signaling pathways of LH/CGR. Moreover, the comparison between the two NAMs on LH/CGR showed ADX68693 being stronger than ADX68692 (*p-value* < 0.001 at 100 nM of hCG), which is consistent with the kinetic data shown in **Fig. 2**.

**3.2. Effects of ADX compounds on β-arrestin 2 recruitment to LH/CGR**

Next, the effect of ADX68692 and ADX68693 on the recruitment of β-arrestin 2 to the activated LH/CGR in real-time and in live cells was investigated using BRET technology. For this, cells co-expressing LH/CGR-Rluc8 (BRET donors) and yPET-ß-arrestin 2 (BRET acceptor) were used as previously reported . As shown in **Fig. 4**, the dose-dependent response of hCG is consistent with the previous study using similar assay . Consistent with gene reporter data shown in **Fig. 3**, increasing doses of both ADX68692 (**Fig. 4A**) and ADX68693 (**Fig. 4B**) strongly reduced hCG-promoted ß-arrestin 2 recruitment, with maximal effect observed at 10 μM of NAMs indicating a negative allosteric effect (**Table 1**). The comparison between the two NAMs at the maximal response of the hormone observed at 100 nM showed that ADX68693 is significantly more potent than ADX68692 (*p-value < 0.01* at 0.5 μM and 1 μM of NAMs,) on LH/CGR (**Fig. 4C**) with pIC$_{50}$ values of 6.02 ± 0.11 and 6.824 ± 0.11 for ADX68692 and ADX68693, respectively. Again, we used the recruitment of ß-arrestin 2 to V2R as a negative control, showing that neither ADX68692 nor ADX68693 affected DDAVP-promoted V2R-ß-arrestin 2 association (**Fig. 4D**) and further demonstrating the specificity of these compounds on gonadotropin receptors.

In addition, real-time kinetics was performed using 10 μM of NAMs and 100 nM of hCG. After 20 minutes of pre-treatment with either ADX68692 or ADX68693, a total inhibition of hCG-promoted ß-arrestin 2 recruitment was observed with both NAMs (**Fig. 4E**). Such an inhibition is consistent with that observed in dose-response analysis (**Fig. 4A** and **B**). In order to examine whether hCG-mediated ß-arrestin 2 response could be reversed with the NAMs, we performed real-time kinetics where cells were first stimulated with hCG to promote ß-arrestin 2 recruitment and 12 minutes later either NAM (10 μM) was applied. Interestingly, the addition of NAMs inhibited ß-arrestin 2 recruitment very rapidly, indicating that the action of the two compounds occurred even when the receptor is fully activated by hCG (**Fig. 4F**) such as in the physiological situation. Together these data demonstrate the profound inhibition of ß-arrestin recruitment to LH/CGR by

ADX68692 and ADX68693.

## 3.3. Effect of ADX compounds on LHR-mediated steroidogenesis in mLTC-1 and primary rat Leydig cells

In order to link the foregoing data on LH/CGR in transfected HEK293 cells to a more integrated physiological response, a murine Leydig tumor cell line (mLTC-1) as well as primary rat Leydig cells, known to endogenously express LHR and to trigger hCG responses were used . Thus, the antagonistic effects of the two NAMs at 10 μM were examined on cAMP, progesterone, and testosterone production in both mLTC-1 and primary rat Leydig cells. In cAMP assay, both ADX68692 and ADX68693 nicely inhibited hCG-promoted cAMP production in mLTC-1 (**Fig. 5A**) and primary rat Leydig cells (**Fig. 5B**) cells. This was observed even at high concentrations of hCG, which is consistent with the NAM action of the two compounds observed in the gene reporter assay in HEK293 cells (**Fig. 3**), demonstrating an effect on hCG Emax as illustrated in (**Table 2**). Furthermore, these data are consistent with cAMP data obtained in HEK293 cells since ADX68693 led to a significantly stronger inhibition of cAMP production than ADX68692 (*p-value < 0.01* for 100 nM of hCG) with a slight shift in the potency of hCG (**Table 2**).

Next, we examined the effects of ADX68692 and ADX68693 on progesterone and testosterone production induced by hCG stimulation in mLTC-1 and primary rat Leydig cells using HTRF®-based assays. For progesterone in mLTC-1 cells, neither ADX68692 nor ADX68693 affected hCG-induced progesterone production (**Fig. 5C**) as illustrated in **Table 2**. By contrast, in primary rat Leydig cells while ADX68692 surprisingly appeared to potentiate (by ~37%) hCG-promoted progesterone, ADX68693 dramatically (by ~70%) inhibited such a response (**Fig. 5D**) (**Table 3**). However, in testosterone assay the data obtained in mLTC-1 (**Fig. 5E**) and primary rat Leydig (**Fig. 5F**) cells are very consistent since ADX68692 only partially inhibited hCG-promoted response whereas ADX68693 had a full antagonistic effect (*p-value < 0.001* when ADX68692 and ADX68693 were compared with each other from 0.1 to 100 nM of hCG). Indeed, the strength of the two NAMs on the maximal testosterone response of hCG in both cell models were to similar levels with ~30-40% and ~90% of inhibition by ADX68692 and ADX68693, respectively (**Table 2** and **3**).

Together, these observations suggest that the two NAMs present biased effects on the LHR in mLTC-1 and primary rat Leydig cells. In addition, the bias calculation revealed that, when compared to DMSO, both ADX68692 (**Fig. 6A**) and ADX68693 (**Fig. 6B**) lead to significant bias toward progesterone production compated to cAMP response, in mLTC-1 cells: i.e. both NAMs are significantly more potent to inhibit cAMP response than progesterone production in mLTC-1 cells (**Table 4**). In addition, to a certain extent ADX68692 also shows a bias toward testosterone compared to cAMP response (**Table 4**). For both NAMs, the inhibition of steroid production in mLTC-1 cells showed in **Fig. 5C** and **E** nicely showed biased effects towards progesterone compared to testosterone as illustrated in **Fig. 6C**, even though the lack of identifiability of the testosterone response (the transduction coefficient is poorly identifiable), due to almost full inhibition by ADX68693, prevents to obtain a significant statistical result (**Table 4**). By contrast, in primary rat Leydig while ADX68692, when compared to DMSO, showed significant bias toward progesterone production compared to both cAMP response and testosterone production, ADX68693, in turn, when compared to DMSO, showed an opposite bias, toward cAMP compared to progesterone production: ADX68693 has a stronger inhibitory effects of the hCG response on progesterone production than cAMP (Table 5).

Finally, the comparison bewen the two NAMs indicated that the inhibitory effects of ADX68693 are moderately biased with respect to cAMP and testosterone responses compared to progesterone production, both in mLTC-1 cells (**Fig. 6C** and **Table 4**) and primary rat Leydig cells (Fig 7C and Table 5).

## 4. Discussion

Two compounds, ADX68692 and ADX68693, have been recently reported to act as NAMs at the FSHR, leading to the inhibition of FSH-promoted steroidogenesis in rat granulosa primary cells and follicle maturation *in vivo* . Interestingly, despite their structural similarities, the two molecules presented different antagonistic profiles at the FSHR. While ADX68692 blocked FSHR-promoted cAMP production and progesterone as well as estradiol production, ADX68693 inhibited cAMP and progesterone with the same efficacy as ADX68692 but did not block estradiol production . Thus, because of structural and signaling similarities between FSHR and LH/CGR, as well as their implication in the physiology of reproduction, the effects of ADX68692 and ADX68693 were studied on LH/CGR. We investigated the effects on the two major transduction mechanisms known to operate at the LH/CGR: Gs/cAMP/PKA and ß-arrestins. Those studies were extended to examine the compounds' effects on the control of steroidogenesis using mLTC-1 cell line and primary rat Leydig cells endogenously expressing LHR. Together, our data clearly demonstrate that both ADX68692 and ADX68693 inhibited LH/CGR activation by hCG in HEK293 and mLTC-1 as well as primary rat Leydig cells. Indeed, both cAMP production and ß-arrestin recruitment induced by hCG were inhibited in a dose-dependent manner. In addition, differential inhibition of hCG-promoted steroid production by ADX68692 and ADX68693 was observed when mLTC-1 cell line and primary rat Leydig cells were compared. These constitute the first small molecules antagonizing LH/CGR since only agonist compounds were reported so far for this receptor . This finding is of great importance regarding the potential application of ADX68692 and ADX68693 to control the reproductive activity, considering the involvement of both FSHR and LH/CGR in this function and their co-expression within granulosa cells at specific stages of the female reproductive cycle. In fact, the physiological effects of FSH and LH on the ovary are characterized by the stimulation of the production of estradiol and progesterone, which play key roles in ovarian function and control of the reproductive cycle . The mechanisms involved in the regulation of progesterone production by ovarian granulosa cells imply the activation of the Gs/cAMP/PKA pathway leading to the modulation of gene expression associated with steroidogenesis such as the steroidogenic acute regulatory protein (StAR), 3ß-hydroxysteroid dehydrogenase (3ßHSD), and the cytochrome P450

(P450scc) enzyme system . Other studies also reported a crosstalk between ERK activation and progesterone production downstream of PKA, with ERK exerting a negative feedback on steroid production . For LH/CGR, other molecular mechanisms, including its transactivation with EGFR were proposed to play role in steroidogenesis and thereby oocyte maturation, and in gonadotropin-stimulated follicles . Taken together with the previous data on the FSHR , our study using ADX68692 and ADX68693 highlights the complexity of the mechanisms involved in the control of steroidogenesis *via* FSHR and LH/CGR.

In cAMP assay, the allosteric antagonism of both NAMs was also shown to reverse gonadotropin-induced cAMP production in real-time kinetic assay (**Fig. 2**). The relatively slow kinetics of such inhibition is likely due to strong accumulation of cAMP before the addition of NAMs since the assay is based on intracellular cAMP using BRET sensor. Moreover, kinetic and dose-response analysis indicated that ADX68693 is more efficient than ADX68692 to inhibit hCG-promoted cAMP production in HEK293 cells, whereas the opposite situation seems to happen on FSHR in HEK293 cells with ADX68692 being more effeicient than ADX68693 (data not shown for the kinetics). These obseravtions suggest different allosteric properties of the two NAMs towards FSHR and LH/CGR, which may be explained by the differences in the specific domains involved in the binding of NAMs on each receptor. Moreover, these different profiles may explain the differential effects of the two NAMs on steroidogenesis whether FSHR or LHR is targeted (see below). From the structure point of view, the only major difference between ADX68692 and ADX68693 is the presence of pyridine group in ADX68692 . This may determine the allosteric properties of the NAMs and thereby explain the differences between FSHR and LH/CGR. The fact that there was no significant difference in NAMs' efficacies on LH/CGR-mediated cAMP response in HEK293 cells is likely due to the high level of receptor expression achieved in this system. Indeed, we recently reported that in a similar experimental setting, maximal cAMP response could be reached with less than 5% of receptor occupied , meaning that a substantial amount of spare receptors are available thereby preventing the NAMs to reduce maximum efficacies and rather leading to shifts in $EC_{50}$ values of hCG . Whereas in mLTC-1 and primary rat Leydig cells, both NAMs profoundly decreased the maximal hCG-induced cAMP production with ADX68693 appearing more efficient than ADX68692 (**Table 2** and **3**). Again, this likely reflects the fact that mLTC-1 and primary rat Leydig cells express less receptors, hence less spare receptors, than transfected HEK293 cells. These data were also consistent with the reporter gene assay indicating downstream inhibition of the cAMP/PKA/CREB-dependent pathway.

In addition to Gs/cAMP pathway, ß-arrestins are known to play a major role not only in the desensitization of GPCRs but also in their ability to signal independently of G proteins . Indeed, many studies demonstrated the interaction of ß-arrestins with gonadotropin receptors and, in some cases, their involvement in receptor signaling . Moreover, the role of ß-arrestins in steroidogenic pathways has been demonstrated for aldosterone production mediated by the angiotensin II receptor (AT1R) . In the present study, we report that, at maximal dose (10 μM), the two NAMs completely abolished ß-arrestin 2 recruitment to LH/CGR. Real-time kinetics showed that both NAMs very rapidly reversed hCG-promoted ß-arrestin 2 recruitment in HEK293 cells. The fact that, in the same cellular system (i.e. transfected HEK293 cells), the NAMs led to a shift in $EC_{50}$ for cAMP response whereas they had profound effects on hCG maximal response when measuring ß-arrestin 2 recruitment is consistent with our previous finding that 100% receptor occupancy needs to be reached in order to achieve maximal ß-arrestin recruitment . This is nicely illustrated by the shift (3 logs) in hCG dose-response when ß-arrestin 2 response was compared to cAMP response.

Interestingly, when we attempted to link our data on cAMP and ß-arrestins observed in HEK293 cells to hCG-promoted steroid production in mLTC-1 and primary rat Leydig cells, our data were reminiscent of the previous study on FSHR . In addition, our data in mLTC-1 taken along with those obtained in primary rat Leydig cells showed the complexity of the pathways linking cAMP pathway and progesterone production. In fact, Dias et al. showed that ADX68692 blocked FSHR-promoted cAMP production and progesterone as well as estradiol production, while ADX68693 acted with the same efficacy as ADX68692 on cAMP and progesterone inhibition but interestingly did not block estradiol production . However, here we clearly showed that neither ADX6892 nor ADX68693 had significant effects on progesterone production in mLTC-1 cells even though both nicely decreased hCG-induced cAMP production. In contrast, in primary rat Leydig cells ADX68693 but not ADX68692 nicely inhbited hCG-promoted progesterone production.

Wethere these differences between mLTC-1 and primary rat Leydig cells are due to the species (mice *versus* rat) or cells (cell line *versus* primary cells) is still unclear. This also true for the unexpected potentiating effects of ADX68692 observed in primary rat Leydig cells compared to mLTC-1 cells.

For testosterone responses, the situation is more concordant between mLTC-1 and primary rat Leydig cells since ADX6892 led to partial inhibition whereas ADX68693 completely abolished hCG-promoted testosterone production. In addition, these data confirm the better efficacy of ADX68693 compared to ADX68692 on LH/CGR. This reveals interesting biased effects of both NAMs on LH/CGR: both NAMs blocked the canonical Gs/cAMP pathway and ADX68692 only partially inhibited hCG-induced testosterone but ADX68693 completely inhibited hCG-induced testosterone production in both mLTC-1 and primary rat Leydig cells. However, neither compounds significantly affected progesterone response in mLTC-1 cells but in primary rat Leydig cells ADX68693 dramatically inhibited progesterone while ADX68692 would have a positive effect. Such biased effects were nicely supported by our bias quantification using the original operational model and recently applied for other GPCRs (**Fig. 6**). The original work on FSHR in rat granulosa primary cells clearly supports the link between cAMP pathway and progesterone production, indicating that alternative and/or additional pathways are involved in estradiol production. In line with this, our data suggest that progesterone and testosterone production induced by LHR in mLTC-1 cells may not entirely depend on the cAMP pathway. Thus, progesterone and testosterone production seems to be controlled by distinct signaling pathways yet to be identified.

The total inhibition by both NAMs of ß-arrestin 2 recruitment to LH/CGR in HEK293 cells suggests that this transduction mechanism may play an important role in testosterone production. Together, these observations further illustrate the complexity of the mechanisms involved in the control of steroidogenesis *via* FSHR and LH/CGR. Such a discrepancy between cAMP/ß-arrestin inhibition and steroid production is in fact difficult to reconcile with the classical view postulating that progesterone production depends on the activation of the Gs/PKA/cAMP pathway. One possibility is the engagement of Gs- and ß-arrestin-independent transduction mechanisms. Candidates include other G proteins (Gq/Gi) known to couple to both FSHR and LH/CGR (for review ) and could be differentially inhibited by the NAMs. Alternatively, the absence of significant inhibition of progesterone in mLTC-1 cells may be due to the residual cAMP response even upon treatment with NAMs. Moreover, our data cannot rule out the possibility of differential inhibition of intermediate pathways downstream of cAMP or ß-arrestins controlling steroid production. This may be consistent with a scenario where the activation of both Gs/cAMP and ß-arrestin pathways would be involved to a different extent (e.g. with different activation thresholds) in the production of both progesterone and testosterone production. Differences in the kinetics of progesterone and testosterone production may also account for the observed biased effects between the two NAMs .

Finally, this study demonstrates that ADX68692 and ADX68693 antagonized LH/CGR with differential profiles regarding their canonical Gs/cAMP and ß-arrestin pathways. Even though their cross-reactivity on LH/CGR has yet to be demonstrated *in vivo*, our findings suggest that these compounds may impact the steroidogenic pathways differently as a function of FSHR and LH/CGR relative expression levels. Exploring this possibility will require further investigations *in vitro* and *in vivo*. It will be important to take into account the fact that the two receptors are co-expressed in the same follicular cells at specific stages of the female reproductive cycle. Interestingly, recent studies in transfected cells reported heterodimerization between FSHR and LH/CGR with the physiological relevance still to be demonstrated . Nevertheless, the intriguing possibility exists that the pharmacological profiles of the two NAMs on FSHR-LH/CGR heterodimer could be different compared to the respective protomers or homodimers. Furthermore, the differences in the biased effects of ADX68692 and ADX68693 with respect to steroid production that was previously observed on FSHR  and that our study revealed on LH/CGR as well, suggest that the combination of both NAMs may be required for efficient contraceptive or therapeutic applications to achieve full inhibition of steroidogenesis.

**Acknowledgments**

This work was funded by ARTE2, MODUPHAC, "ARD2020 Biomédicament" grants from Région Centre. MAA is funded by LE STUDIUM® Loire Valley Institute for Advanced Studies and AgreenSkills.

Table 1: Efficacy (*Emax*) and efficiency (*pEC$_{50}$*) of hCG on LH/CGR measured in different assays in HEK293 cells treated or not with 10 µM of NAMs.

| Treatments / responses | hCG + DMSO | | hCG + ADX68692 | | hCG + ADX68693 | |
|---|---|---|---|---|---|---|
| | *Emax (%)* | *pEC$_{50}$* | *Emax (%)* | *pEC$_{50}$* | *Emax (%)* | *pEC$_{50}$* |
| cAMP | 100 | 10.67 ± 0.05 | 92.01 ± 2.97 | 9.66 ± 0.09 | 105 ± 2.27 | 8.93 ± 0.06 |
| pSOM-Luc | 100 | 9.35 ± 0.09 | 55.89 ± 4.40 | 8.78 ± 0.19 | 31.57 ± 2.30 | 8.88 ± 0.17 |
| ß-arrestin 2 | 100 | 8.12 ± 0.08 | 10.65 ± 1.62 | ND | 4.99 ± 2.15 | ND |

Table 2: Efficacy (*Emax*) and efficiency (*pEC$_{50}$*) of hCG on LHR measured in different assays in mLTC-1 cells treated or not with 10 µM of NAMs.

| Treatments / responses | hCG + DMSO | | hCG + ADX68692 | | hCG + ADX68693 | |
|---|---|---|---|---|---|---|
| | *Emax (%)* | *pEC$_{50}$* | *Emax (%)* | *pEC$_{50}$* | *Emax (%)* | *pEC$_{50}$* |
| cAMP | 100 | 10.10 ± 0.09 | 42.67 ± 2.53 | 9.16 ± 0.18 | 23.80 ± 2.71 | 9.14 ± 0.34 |
| Progesterone | 100 | 11.28 ± 0.14 | 99.84 ± 10.14 | 11.13 ± 0.39 | 95.29 ± 8.22 | 11.32 ± 0.35 |
| Testosterone | 100 | 11.34 ± 0.18 | 67.73 ± 4.22 | 10.77 ± 0.21 | 7.94 ± 2.23 | ND |

Table 3: Efficacy (*Emax*) and efficiency (*pEC$_{50}$*) of hCG on LHR measured in different assays in primary rat Leydig cells treated or not with 10 µM of NAMs.

| Treatments / responses | hCG + DMSO | | hCG + ADX68692 | | hCG + ADX68693 | |
|---|---|---|---|---|---|---|
| | *Emax (%)* | *pEC$_{50}$* | *Emax (%)* | *pEC$_{50}$* | *Emax (%)* | *pEC$_{50}$* |
| cAMP | 100 | 10.01 ± 0.06 | 63.85 ± 3.05 | 9.52 ± 0.14 | 33.66 ± 3.01 | 9.56 ± 0.28 |
| Progesterone | 100 | 12.19 ± 0.22 | 136.80 ± 3.78 | 11.77 ± 0.12 | 30.38 ± 6.34 | 10.70 ± 1.05 |
| Testosterone | 100 | 11.59 ± 0.33 | 61.89 ± 4.11 | 11.44 ± 0.25 | 11.93 ± 5.11 | ND |

Table 4: Bias factor of the effects of both NAMs on hCG-promoted responses in mLTC-1 cells calculated as described *Material and Methods*.

| Treatments / responses | ADX68692 / DMSO | | ADX68693 / DMSO | | ADX68693 / ADX68692 | |
|---|---|---|---|---|---|---|
| | Bias factor | *p*-value | Bias factor | *p*-value | Bias factor | *p*-value |
| Progesterone/cAMP | 8.24 | 0.012 | 19.27 | 0.033 | 2.34 | 0.381 |
| Testosterone/cAMP | 2.49 | 0.123 | 0.94 | 0.979 | 0.38 | 0.652 |
| Progesterone/testosterone | 3.31 | 0.024 | 20.46 | 0.186 | 6.18 | 0.404 |

**Table 5: Bias factor of the effects of both NAMs on hCG-promoted responses in primary rat Leydig cells calculated as described *Material and Methods*.**

| Treatments / responses | ADX68692 / DMSO | | ADX68693 / DMSO | | ADX68693 / ADX68692 | |
|---|---|---|---|---|---|---|
| | **Bias factor** | *p*-value | **Bias factor** | *p*-value | **Bias factor** | *p*-value |
| **Progesterone/cAMP** | 5.88 | 0.0002 | 0.24 | 0.09 | 0.04 | 0.004 |
| **Testosterone/cAMP** | 0.47 | 0.32 | 13.61 | 0.74 | 29.06 | 0.67 |
| **Progesterone/testosterone** | 12.55 | 0.03 | 0.02 | 0.62 | 0.001 | 0.44 |

**Figure Legends**

**Fig. 1: Dose-response analysis of the effects of ADX compounds on cAMP production.** HEK293 cells stably expressing FSHR (**A**) or transiently co-expressing LH/CGR (**B**, **C**, and **D**) or V2R (**E**) with cAMP pGloSensor™-22F (**A**) or the cAMP-BRET sensor (**B**, **C**, **D**, and **E**), were used for dose-response analysis of hormone-promoted cAMP production. For this, cells were first pretreated or not for 20 minutes at 37°C with the different concentrations of ADX68692 or ADX68693 as indicated. Then, cells were stimulated or not for 30 minutes at 37°C with increasing concentrations of FSH (**A**), hCG (**B** and **C**) or DDAVP (**E**) before luminescence and BRET measurements. The curves in panel **D** were generated by pretreating cells with increasing concentrations of NAMs followed by stimulation with 0.1 nM of hCG. Data are means ± SEM of three experiments performed in duplicate.

**Fig. 2: Real-time kinetics of the effects of ADX compounds on cAMP production.** HEK293 cells transiently co-expressing LH/CGR and the cAMP-BRET sensor were used for kinetic analysis of hCG-promoted cAMP production. For this, cells were first pretreated (**A**) or not (**B**) for 20 minutes at 37°C with either DMSO or 10 μM of ADX68692 or ADX68693 as indicated. Then, BRET measurements were rapidly performed upon addition of 1 nM of hCG followed (**B**) or not (**A**) by NAMs addition as indicated. Data are representative of three experiments performed in triplicate.

**Fig. 3: Effects of ADX compounds on cAMP sensitive reporter gene expression.** HEK293 cells transiently co-expressing the cAMP sensitive reporter gene (pSOM-Luc) and either the FSHR (**A**) or LH/CGR (**B**) were first starved overnight and pretreated or not for 1 hour at 37°C with 10 μM of ADX68692 or ADX68693. Then cells were stimulated or not for 6 hours at 37°C with increasing concentrations of FSH (**A**) or hCG (**B**) before luciferase luminescence was measured using Bright-

Glo™ luciferase assay. Data are means ± SEM of four independent experiments performed in single point.

**Fig. 4: Effects of ADX compounds on ß-arrestin 2 recruitment**. HEK293 cells transiently co-expressing yPET-ß-arrestin 2 and either LH/CGR-Rluc8 (**A**, **B**, **C**, **E** and **F**) or V2R-Rluc8 (**D**), were used for dose-response and real-time kinetic analysis of hormone-promoted BRET increases. For dose-responses, cells were first pretreated or not for 20 minutes at 37°C with the different concentrations of ADX68692 or ADX68693 as indicated. Then, cells were stimulated or not for 30 minutes at 37°C with the increasing concentrations of hCG or DDAVP before BRET measurements were performed. The curves in panel **C** were generated by pretreating cells with increasing concentrations of NAMs followed by stimulation with 100 nM of hCG. For the kinetics, cells were first pretreated (**A**) or not (**B**) for 20 minutes at 37°C with either DMSO or 10 μM of ADX68692 or ADX68693 as indicated. Then, BRET measurements were rapidly performed upon addition of 100 nM of hCG followed (**B**) or not (**A**) by NAMs addition as indicated. Data are means ± SEM of three-six experiments performed in duplicate.

**Fig. 5: Effects of ADX compounds on LHR activation in mLTC-1 and primary rat Leydig cells.** mLTC-1 (**A**, **C**, and **E**) and primary rat Leydig (**B**, **D**, and **F**) cells endogenously expressing the LHR were used for cAMP, progesterone and testosterone production as indicated. For cAMP, cells were first pretreated for 20 minutes at 37°C with either DMSO 10 μM of ADX68692 or ADX68693, and then stimulated or not for 30 minutes at 37°C with the increasing concentrations of hCG before cAMP production was assessed by HTRF®-based assay, as described in Material and Methods. For steroid production, cells were first pretreated for 1 hour at 37°C with either DMSO or 10 μM of ADX68692 or ADX68693 as indicated. Then, cells were stimulated or not for 3 hours (for primary rat Leydig cells) or 24 hours (for mLTC-1 cells) at 37°C with increasing concentrations of hCG before progesterone (**C** and **D**) and testosterone (**E** and **F**) production was quantified using HTRF®-based assay. Data are means ± SEM of three independent experiments performed in duplicate (for mLTC-1 cells) or six independent experiments performed in single point (for Leydig cells).

**Fig. 6: Bias plots of NAMs effects in mLTC-1 cells.** The biased effects of both NAMs on hCG-promoted responses in mLTC-1 cells were calculated as described Material and Methods. The analysis of hCG-induced cAMP, progesterone, and testosterone was performed at different doses of hCG and this for the different treatments, DMSO *versus* ADX68692 (**A**), DMSO *versus* ADX68693 (**B**), and ADX68692 *versus* ADX68693 (**C**) as indicated.